\documentclass[epsfig,12pt] {article}
\usepackage{graphics}
\usepackage{epsfig}
\usepackage{amssymb}
\usepackage{amsmath}
\usepackage{bm}
\usepackage{epsfig}

\def\n{\noindent}

\begin{document}

\baselineskip .7cm

\author{ Navin Khaneja \thanks{To whom correspondence may be addressed. Email:navinkhaneja@gmail.com} \thanks{Department of Systems and Control Engineering, IIT Bombay, Powai - 400076, India.}}

\vskip 4em

\title{\bf Conservation of energy, density of states and spin lattice relaxation}

\maketitle

\vskip 3cm

\begin{center} {\bf Abstract} \end{center}
The starting point of all NMR experiments is a spin polarization which develops when we place the sample in static magnetic field $B_0$. There are excess of spins aligned along $B_0$ (spin up with lower energy) than spins aligned opposite (spin down with higher energy) to the field $B_0$. A natural question is what is the source of this excess spin polarization because relaxation mechanisms can flip a up spin to a down spin and vice-versa. The answer lies in the density of states. When a molecule with spin down flips to spin up it loses energy. This energy goes into increasing the kinetic energy of the molecule in the gas/solution phase. At this increased kinetic energy, there are more rotational-translational states accessible to the molecule than at lower energy. This increases the probability the molecule will spend in spin up state (higher kinetic energy state). This is the source of excess polarization. In this paper, we use an argument based on equipartition of energy to explicitly count the excess states that become accessible to the molecule when its spin is flipped from down to up. Using this counting, we derive the familiar Boltzmann distribution of the ratio of up vs down spins. Although prima facie, there is nothing new in this paper, we find the mode counting argument for excess states interesting. Furthermore, the paper stresses the fact that spin polarization arises from higher density of states at increased kinetic energy of molecules.

\vskip 3em

\section{Introduction}

Modern NMR experiments use a large static magnetic field $B_0$ of order of $10-20$ tesla to create a population difference between spin up and down states. If $n_{\uparrow}$ and $n_{\downarrow}$ denote the number of spin up and down spins in a population then $\frac{n_{\uparrow}}{n_{\downarrow}} = \exp(\frac{\Delta E}{k_BT})$ where
$\Delta E = \hbar \omega_0$ is the energy difference between spin up and down states and $\omega_0$ is the Larmor frequency of the spins in $B_0$ given by $\omega_0 = \gamma B_0$, where $\gamma$ 
is the gyromagnetic ratio of the spins. Then $\frac{n_{\uparrow} -n_{\downarrow}}{n_{\downarrow}} = \frac{\Delta n}{n_{\downarrow}}\sim \frac{\hbar \omega_0}{k_BT} $ is the excess population. For protons at $B_0 = 14$ tesla, $\omega_0 = 600$ MHz, which gives $\hbar \omega_0 \sim 10^{-25}$ where as at room temperature $k_BT \sim 10^{-20}$ giving   $\frac{\Delta n}{n_{\downarrow}} \sim 10^{-5}$. Thus at room temperature only one out of $10^5$ spins is preferentially in up state. For example, in a sample with $10^{19}$ spins, we will have excess of $10^{14}$ up spins.  This gives the sample a net magnetization $M$ which is rotated and flipped with rf-pulses in an NMR experiment. After the magnetization is flipped, there are excess of down spins. After a recovery time termed $T_1$, the magnetization recovers to its original equilibrium distribution by phenomenon of spin lattice relaxation \cite{abragam}. Spin lattice relaxation flips excess spins back to up position.

\n In gas/solution phase, this relaxation arises due to stochastic tumbling of molecules, which arises due to random collisions and rotation of molecules. Without going too much into details, we mention two relaxation
mechanisms. The field $B_0$ at site of nuclear spin is shielded by electronic currents which are described by a tensor. When molecules tumbles, this tensor reorients, causing a fluctuating field at the nuclear spin site. This relaxation mechanism is called chemical shift anisotropy \cite{cavanagh, goldman} and the resulting fluctuating field can flip a spin.  
If there are two spins on a molecule that are coupled by a dipolar interaction, then the dipolar field seen by a spin due to neighbouring spin depends on the orientation of internuclear vector. When the molecule tumbles, this orientation changes and dipolar field fluctuates and can result in a spin flip. This is called dipolar relaxation \cite{cavanagh, goldman}.

Suffice to say, there are mechanisms for spin to flip form down to up and vice versa. However this begs the question, when a spin can flip both ways due to relaxation why do we see excess of up spins. The answer lies in the density of states. When a molecule with spin down flips to spin up it loses energy. This energy goes into increasing the kinetic energy of the molecule in the gas/solution phase. At this increased kinetic energy, there are more rotational-translational states accessible to the molecule. This increases the probability the molecule will spend in spin up state (higher kinetic energy state). This is the source of excess polarization. We now make an argument based on equipartition of energy to count the excess states that become accessible to the molecule when its spin is flipped from down to up. Using this counting, we derive the familiar Boltzmann distribution of the ratio of up vs down spins.

\section{Theory}

Let us consider molecules of a gas carrying spin. The kinetic energy of the molecule has two components, the translational velocity and rotational velocity. We consider particle in a box model. Consider gas molecules of mass $M$ in a rectangular box of length $L$ and volume $V = L^3$. The orbitals of the gas molecule have the form \cite{kittel}

\begin{equation}
\psi(x, y,z) \propto \sin(\frac{\pi n_x x}{L})\sin(\frac{\pi n_y y}{L})\sin(\frac{\pi n_z z}{L}).
\end{equation} where  $(n_x, n_y, n_z)$ are nonnegative integers.

The energy

\begin{equation}
E_t = \frac{\hbar^2 \pi^2}{2ML^2} (n_x^2 + n_y^2 + n_z^2).
\end{equation} The translational modes are enumerated by quantum numbers $(n_x, n_y, n_z)$, with

\begin{equation}
\label{eq:enum}
n_x^2 + n_y^2 + n_z^2 = \frac{2ML^2 E_t}{\hbar^2 \pi^2} = R^2.
\end{equation}

\begin{figure}
  \centering
  \includegraphics[scale=.5]{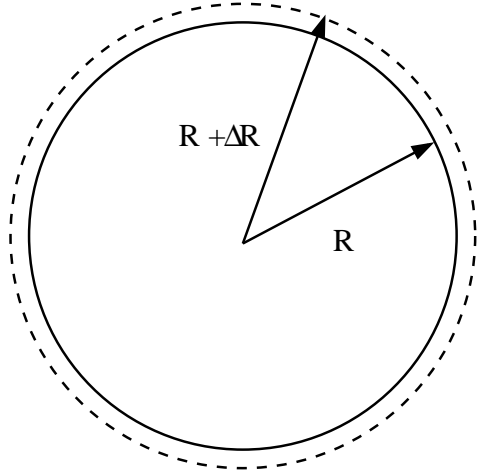}
  \caption{Fig shows number of states enclosed in volume between sphere of radius $R$ and $R + \Delta R$ in Eq. (\ref{eq:enum})} \label{fig:sphere}
\end{figure}

The number of solutions $N$ to Eq. (\ref{eq:enum}) is simply counted by looking at the volume enclosed in sphere
of radius $R$ and $R+ \Delta R$ as shown in figure \ref{fig:sphere}, which is simply $4 \pi R^2 \Delta R \propto E_t$.

At temperature $T$, equipartition of energy gives that energy in the translational modes is $E_t = \frac{3 k_B T}{2}$. When the spin of the molecule flips from down to up state an energy $\hbar \omega_0$ is released. This increases the kinetic energy from $E_t$ to $E_t + \hbar \omega_0$ and hence number of modes increase from $N$ to $ N + \Delta N$. The ratio $\frac{N + \Delta N}{N} = \frac{E_t + \hbar \omega_0}{E_t} = 1 + \frac{2 \hbar \omega_0}{3 k_B T}$ or $\frac{\Delta N}{N} = \frac{2 \hbar \omega_0}{3 k_B T}$.

We now come to rotational energy. If $l$ is the angular momentum quantum number then the rotational kinetic energy is
\begin{equation}
  \label{eq:rotkin}
  E_r = \frac{\hbar^2 l(l+1)}{2 I},
\end{equation}
where $I$ is moment of inertia of the molecule. For a given $l$, there are $L = 2 l + 1$ states given by quantum number $m$, corresponding to angular momentum along say $z$ direction, where $m$ goes in steps of $1$ and $-\l \leq m \leq l$. At temperature $T$, equipartition of energy gives that that energy in rotational modes is $E_r = \frac{3 k_B T}{2}$. When the spin of the molecule flips from down to up state an energy $\hbar \omega_0$ is released. This increases the kinetic energy from $E_r$ to $E_r + \hbar \omega_0$ and hence number of modes increase from $L$ to $ L + \Delta L$. From Eq. \ref{eq:rotkin}, we have $\frac{\Delta E_r}{2E_r} = \frac{\Delta l}{l} = \frac{\Delta L}{L}$. With $\Delta E_r = \hbar \omega_0$, we have  $\frac{\Delta L}{L}=  \frac{\hbar \omega_0}{3 k_B T}$. 

The total number of rotational and translational states are $NL$. They increase by $\Delta N L + N \Delta L$ as the molecule flips from down to up sate. Then

\begin{equation}
  \frac{\Delta N L + N \Delta L}{NL} = \frac{\hbar \omega_0}{k_B T}.
 \end{equation} 

Thus when the molecule loses energy to the bath as a result of  spin flip from down to up state, more states become accessible and we recover the Boltzmann distribution.  Of increase in the number of states, $\frac{1}{3}$ is accounted by rotational and $\frac{2}{3}$ by translational states. In short, as more states are accessible when we are in spin up state there are excess of spin up states.

\section{Conclusion}

We presented a simplified model, that helps us enumerate the increase in the rotational-translational states of a molecule when its spin is flipped from higher energy state to lower energy state. This helped us derive the Boltzmann distribution. Of-course, one can make a standard thermodynamics argument \cite{kittel} that if number of accessible bath states at energy $U_0$ are $\exp(\sigma(U_0))$ and we add additional energy of $\hbar \omega_0$ due to spin flip, the number of states increase to $\exp(\sigma(U_0 + \hbar \omega_0))$ with the ratio $$ \frac{\exp(\sigma(U_0 + \hbar \omega_0))}{ \exp(\sigma(U_0))} =  \frac{\exp(\sigma(U_0) + \frac{d \sigma}{dU_0} \hbar \omega_0))}{ \exp(\sigma(U_0))} = \exp(\frac{\hbar \omega_0}{k_BT}), $$ where $\frac{d \sigma}{dU_0} = \frac{1}{k_B T}$. We instead carried out an explicit counting of states. In our treatment, using equipartition of energy, we could separately account for increase in translational and rotational states when we flip a spin and delineate there contribution to Boltzmann distribution. 

In this paper we treated spins in gas/solution phase where energy released due to spin flip goes into increasing the kinetic energy of the molecules. In solid phase, the molecules are rigid, this excess energy of spin flip increases the number of phonons in a lattice \cite{abragam}. In subsequent work, we will present an explicit mode counting argument for phonons in solids.

\end{document}